\documentclass[11pt]{article}
\usepackage{cite}
\usepackage[a4paper, top=2cm, bottom=2cm, left=1.8cm, right=1.8cm, bindingoffset=0mm]{geometry}
\usepackage{amsmath}
\usepackage{amssymb}
\usepackage{amsfonts}
\usepackage{latexsym}
\usepackage{booktabs,graphicx,hyperref,epsfig}
\allowdisplaybreaks
\def\bea{\begin{eqnarray}}
\def\eea{\end{eqnarray}}
\def\beq{\begin{equation}}
\def\eeq{\end{equation}}
\def\ba{\begin{eqnarray}}
\def\ea{\end{eqnarray}}
\def\be{\begin{equation}}
\def\ee{\end{equation}}

\renewcommand{\S}{\mathcal{S}}

\newcommand{\Ord}{\mathcal{O}}

\newcommand{\sss}{\scriptscriptstyle\rm}
\newcommand{\as}{\alpha_{\sss S}}

\def\({\left(}
\def\){\right)}
\def\[{\left[}
\def\]{\right]}
\def\ab{\bar\alpha}
\def\Cscet{C_{\rm\scriptscriptstyle SCET}}
\def\Cqcd{C_{\rm\scriptscriptstyle QCD}}
\def\sigscet{\sigma_{\rm\scriptscriptstyle SCET}}
\def\sigqcd{\sigma_{\rm\scriptscriptstyle QCD}}

\def    \hepph  #1 {{\tt hep-ph/#1}}
\def    \hepex  #1 {{\tt hep-ex/#1}}
\long\def\symbolfootnote[#1]#2{\begingroup%
\def\thefootnote{\fnsymbol{footnote}}\footnote[#1]{#2}\endgroup}

\numberwithin{equation}{section}
\begin{document}
\begin{flushright}
DESY 12-007\\
IFUM-986-FT
\end{flushright}
\begin{center}
{\Large\bf Threshold resummation in  SCET vs. perturbative QCD:\\
an analytic comparison}
\vspace{0.6cm}

Marco Bonvini,$^{1,2\,*}$ Stefano~Forte,$^{3}$ 
Margherita Ghezzi$^{3,4}$\symbolfootnote[1]{Current address} 
and Giovanni~Ridolfi$^{1}$
\\
\vspace{1.cm}
{\it ~$^1$ Dipartimento di Fisica, Universit\`a di Genova and
INFN, Sezione di Genova,\\
Via Dodecaneso 33, I-16146 Genova, Italy\\ 
~$^2$ Deutsches Elektronen-Synchroton, DESY,\\
Notkestra{\ss}e 85, D-22603 Hamburg, Germany\\
~$^3$Dipartimento di Fisica, Universit\`a di Milano and
INFN, Sezione di Milano,\\
Via Celoria 16, I-20133 Milano, Italy\\
~$^4$Dipartimento di Fisica, Sapienza Universit\`a di Roma and
INFN, Sezione di Roma,\\
Piazzale Aldo Moro 2, I-00185 Roma, Italy} 
\end{center}

\vspace{0.5cm}

\begin{center}
{\bf \large Abstract:}\end{center}

We compare threshold resummation in QCD, as performed using
soft-collinear effective theory (SCET) in the Becher-Neubert approach,
to the standard perturbative QCD formalism based on factorization and
resummation of Mellin moments of partonic cross-sections.  We consider
various forms of the SCET result, which correspond to different
choices of the soft scale $\mu_s$ that characterizes this approach. We
derive a master formula that relates the SCET resummation to the QCD
result for any choice of $\mu_s$. We then use it first, to show that
if SCET resummation is performed in $N$-Mellin moment space by
suitable choice of $\mu_s$ it is equivalent to the standard
perturbative approach.  Next, we show that if SCET resummation is
performed by choosing for $\mu_s$ a partonic momentum variable, the
perturbative result for partonic resummed cross-sections is again
reproduced, but like its standard perturbative counterpart it is beset
by divergent behaviour at the endpoint.  Finally, using the master
formula we show that when $\mu_s$ is chosen as a hadronic momentum
variable the SCET and standard approach are related through a
multiplicative (convolutive) factor, which contains the dependence on
the Landau pole and associated divergence. This factor depends on the
luminosity in a non-universal way; it lowers by one power of log the
accuracy of the resummed result, but it is otherwise subleading if one
assumes the luminosity not to contain logarithmically enhanced
terms. Therefore, the SCET approach can be turned into a prescription
to remove the Landau pole from the perturbative result, but the price
to pay for this is the reduction by one logarithmic power of the
accuracy at each order and the need to make assumptions on the parton
luminosity.
\clearpage

\setcounter{page}{1} \pagestyle{plain}
\section{Threshold resummation and the Landau pole}
\label{sec:intro}

The interest in the resummation of logarithmically enhanced
contributions due to soft gluon radiation in perturbative QCD
(threshold resummation, henceforth) has been recently revived due to
its relevance for many LHC processes, such as
Higgs~\cite{deFlorian:2009hc} or top~\cite{Cacciari:2011hy}
production. Threshold resummation was originally performed (to
next-to-leading logarithmic accuracy) by factorizing the hadronic
cross-section in Mellin space in terms of a luminosity and a partonic
cross-section, and then exponentiating logarithmically enhanced
corrections to the latter to all orders through
eikonal~\cite{Catani:1989ne} or factorization~\cite{Sterman:1986aj}
techniques. Subsequent derivations and generalizations to all
logarithmic orders were obtained, among others, from a suitable
two-scale generalized factorization theorem~\cite{Contopanagos:1996nh}
and through renormalization-group improvement of the kinematics of the
gluon radiation phase-space~\cite{Forte:2002ni}, with an additional
hypothesis of factorization of virtual corrections.

In all these approaches, resummation is performed after Mellin
transformation of the hadronic cross-section, which factorizes it into
the product of a parton luminosity and a partonic cross-section. More
importantly, in Mellin space the partonic cross-section in the soft
limit can be obtained by exponentiating single-particle emission
cross-sections, thanks to the fact that in Mellin space the
$n$-particle longitudinal phase space factorizes.  The large logs
which are resummed are then logarithms of $N$, the variable which is
Mellin conjugate to $\tau$ (a dimensionless ratio which equals one at
threshold), rather than the original $\ln(1-\tau)$.

More recently, factorization and exponentiation were directly
performed at the level of Feynman diagrams, without the need for a
Mellin transformation, using path-integral methods to separate off
soft gluon modes~\cite{Laenen:2008gt,Laenen:2010uz}. In the latter
approach the standard resummed results are readily recovered, but the
way the terms which dominate in the eikonal limit emerge
order-by-order in perturbation theory (and the next-to-eikonal
corrections to them) is particularly transparent. Indeed, an important
use of resummed results is to provide predictions for higher order
terms which can even be used to construct approximate expressions for
unknown fixed-order corrections (see e.g. Ref.~\cite{Moch:2005ky}).

However, regardless of how resummation is proven, resummed expressions
for partonic cross-sections with a fixed logarithmic accuracy in
momentum space (i.e.\ next$^k$-to-leading $\ln(1-z)$, where $z$ is a
partonic scaling variable) turn out to be ill-defined: they lead to
divergent hadronic cross-sections upon convolution with a parton
luminosity~\cite{Catani:1996yz,Forte:2002ni}. This behaviour is
already present at the fixed-coupling level~\cite{Catani:1996yz}, and
it persists when the coupling runs~\cite{Forte:2002ni}.  It can be
traced~\cite{Catani:1996yz} to the fact that the truncation of
resummed results to any finite logarithmic accuracy in momentum space
induces terms which violate longitudinal momentum conservation,
thereby leading to factorial divergence of the perturbative expansion:
the result is well-defined provided only the truncation to finite
logarithmic accuracy is performed in Mellin space (i.e.,
next$^k$-to-leading $\ln N$, rather than next$^k$-to-leading
$\ln(1-z)$), and the Mellin transform is inverted exactly to power
accuracy, i.e.\ retaining terms to all logarithmic orders in $1-z$ and
only neglecting terms which are down by powers of
$1-z$~\cite{Forte:2002ni}. As a consequence, perturbative QCD
resummation, even if derived using a momentum-space argument, must be
performed in Mellin space (to finite logarithmic accuracy) if it is to
respect momentum conservation, and to lead to finite physical
(hadronic) cross-sections.

At the running-coupling level, however, a new difficulty arises:
namely, it turns out that the next$^k$-to-leading $\ln N$ series of
contributions to the partonic cross-section at any finite logarithmic
accuracy, viewed as a series in the strong coupling $\alpha_s$,
corresponds, upon inverse Mellin transformation, to a divergent series
of contributions to the partonic cross-section. This divergence can be
traced to the Landau pole in the strong coupling: as long
known~\cite{Amati:1980ch}, resummed results correspond to effectively
replacing the hard scale $M^2$ at which the strong coupling is
evaluated with a scale $M^2(1-z)^a$ related to the soft-gluon
radiation process (with $a$ a process-dependent exponent, e.g. $a=2$
for Drell-Yan). Because the hadronic observable is found by
convoluting the partonic cross-section with a luminosity , the
integration over parton momenta always intercepts the region $z\to 1$
where the strong coupling blows up, and this manifests itself as a
divergence of the expansion in powers of $\alpha_s(M^2)$. This
divergence, which is of non-perturbative origin, can be removed by
addition of subleading terms: within the commonly used ``minimal
prescription'' of Ref.~\cite{Catani:1996yz} this is done by choosing a
particular integration path to perform the Mellin inversion integral,
which corresponds to adding a term which is more suppressed than any
power of $1/M^2$, while with the more recent ``Borel
prescription''~\cite{Forte:2006mi,Abbate:2007qv} this is done by
adding a higher twist term to make the divergent series Borel
summable.

An alternative approach to resummation can be based on the
soft-collinear effective field theory
(SCET)~\cite{Bauer:2000ew,Bauer:2000yr,Bauer:2001ct,Bauer:2001yt},
which provides~\cite{Bauer:2002nz} an alternative derivation of QCD
factorization: threshold resummation based on SCET was performed in
Refs.~\cite{Manohar:2003vb,Pecjak:2005uh,Chay:2005rz,Idilbi:2005ky,Becher:2006nr}.
This approach provides a powerful alternative way of determining
resummed results for hadronic observables, which can then be used for
phenomenology through the standard Mellin-space formalism of
Ref.~\cite{Catani:1996yz}. However, it was pointed out in
Ref.~\cite{Becher:2006nr} that, thanks to the fact that the effective
theory deals with the hadronic degrees of freedom, in a SCET approach
resummed expressions can be directly derived in terms of the hadronic
kinematic variable, i.e., in practice, SCET allows one to perform the
resummation of $\ln(1-\tau)$, where $\tau$ is a measurable
dimensionless kinematic ratio.  The advantage is that the divergences
related to the need to integrate over the parton kinematics are no
longer present: hence, in particular, the difficulties related to the
Landau pole of the strong coupling disappear.  The approach of
Ref.~\cite{Becher:2006nr} has been subsequently developed for
phenomenology, and applied to various physical processes, such as
deep-inelastic scattering~\cite{Becher:2006mr},
Drell-Yan~\cite{Becher:2007ty} and Higgs~\cite{Ahrens:2008qu}
production.

Henceforth, for brevity, we will refer to the approach of
Ref.~\cite{Becher:2006nr} as SCET approach, and that of
Refs.~\cite{Catani:1989ne,Sterman:1986aj,Contopanagos:1996nh,Forte:2002ni}
as QCD approach. It should be observed, however, that whereas the QCD
implementation of resummation is unique to any finite perturbative
order, the
aforementioned~\cite{Manohar:2003vb,Pecjak:2005uh,Chay:2005rz,Idilbi:2005ky,Becher:2006nr}
alternative implementations of threshold resummation in SCET, to the
best of our understanding, lead to results which differ (possibly by
subleading terms) even when truncated to finite perturbative
order. Here we will concentrate on the SCET approach of
Ref.~\cite{Becher:2006nr}, which has been widely used in particular
for phenomenological applications.

However, results obtained in the approach of Ref.~\cite{Becher:2006nr}
are not easily compared to those obtained using the standard approach
of Refs.~\cite{Catani:1989ne,Sterman:1986aj,Contopanagos:1996nh,Forte:2002ni},
because the direct connection to factorization and resummation at the
level of partonic cross-sections is lost. Indeed, as mentioned, the
presence of the Landau pole implies that the expansion in powers of
$\alpha_s(M^2)$ of the resummed partonic cross-section
diverges. Hence, if the resummed SCET result of
Ref.~\cite{Becher:2006nr} is free of divergences, its expansion to
fixed order must necessarily differ from that of the standard
Mellin-space resummation.

This difference has never been determined so far: its computation is
the goal of this paper. Clearly, its knowledge is crucial in order to
determine the theoretical and phenomenological viability of the SCET
resummation of Ref.~\cite{Becher:2006nr}.  Some phenomenological
comparisons of resummed predictions for relevant physical processes
obtained using SCET to standard perturbative results have been
performed in
Refs.~\cite{Becher:2006mr,Becher:2007ty,Ahrens:2008qu}. Differences
are found to be reasonably small: however, this does not shed light on
their analytic form. But knowledge of this analytic form is necessary
if we wish to know, first, whether up to the stated accuracy the SCET
and QCD approaches are equivalent, and second, even if they are, what
is the kind of subleading suppression of the terms introduced in the
SCET approach to tame the perturbative divergence, i.e.\ what is the
accuracy of the SCET approach (be it power or logarithmic).

The answer to these questions is presented here in several steps.  In
Section~\ref{sec:thres}, after summarizing the known form of resummed
results both in the perturbative QCD and SCET approach, we recall that
the definition of next$^k$-to-leading log accuracy in the SCET
approach of Ref.~\cite{Becher:2006nr} and in the standard perturbative
QCD approach are different, and only agree at the leading log
level. Beyond the leading log, SCET results are always less accurate
by one power of log than the perturbative ones: so N$^k$LL in the
perturbative case always include terms which only appear in the
N$^{k+1}$LL SCET result, and so forth.  In order to proceed to a
comparison, it is necessary to discuss the dependence of the SCET
resummation on the soft scale: in Section~\ref{sec:soft} we summarize
how SCET results in Mellin space, or in momentum space, at either the
partonic or hadronic level can be obtained by different choices of
soft scale.  A comparison is then made possible through the derivation
of a general relation between the SCET result and the standard result,
by expressing the latter in terms of the convolution of the former
with a function $C_r$ which depends on the soft scale. We establish
this result at next-to-next-to-leading logarithmic order (but we
conjecture it to hold to all orders): it provides a master formula
which enables a full comparison of the QCD and SCET results, both from
an analytic and a numerical point of view.

As a preliminary step, this master formula can be used to prove the
fact that if SCET resummation is performed in Mellin space it is
completely equivalent to the standard approach, and in particular it
has the same logarithmic accuracy at each order. This result was
established already in Refs.~\cite{Becher:2006mr,Becher:2007ty}, but
with the aforementioned lower log accuracy of the SCET results.  This
is done in Section~\ref{sec:partonic}, where we also digress to show
that if SCET resummation is performed in momentum space by choosing a
partonic scaling variable $z$, it coincides with the perturbative
result up to power suppressed corrections, but, like the perturbative
result, it diverges at the partonic endpoint $z=1$.  We can then (in
Section~\ref{sec:hadronic}) tackle the computation of the function
$C_r$ which relates the SCET and perturbative resummation when the
soft scale is chosen as a measurable hadronic scale. In this case, the
SCET result is free of Landau pole, and thus the divergence is
entirely contained in the $C_r$ function. This function depends on the
PDF luminosity in a non-universal way, and thus whether or not it is
subleading depends on the form of the PDF.  In particular, if one
assumes that the luminosity does not contain logarithmically enhanced
terms, then we can show that this function is always logarithmically
subleading, provided only the less accurate SCET definition of
logarithmic accuracy is used. However, any logarithmically enhanced
contribution to the parton luminosity ${\cal L}(x)$ proportional to $
\ln^k(1-x)$, with $k\ge1$, will lead to contributions to $C_r$ which
are of the same order as those induced by perturbative resummation.

Therefore, we conclude that it is only for a particular class of
luminosities that SCET with a hadronic choice of soft scale reproduces
the perturbative result, and can thus be considered to be equivalent
to the standard approach and to provide an alternative prescription to
remove the divergence of the perturbative expansion. Even when this is
the case, the momentum-space SCET resummation prescription of
Ref.~\cite{Becher:2006nr} requires lowering by one order (one power of
log) the accuracy of the resummed result at each logarithmic
order. Furthermore, in the SCET prescription, terms which are
introduced in order to remove the perturbative divergence are only
logarithmically subleading, rather than being power suppressed (as in
the Borel prescription) or exponentially suppressed (as in the minimal
prescription), along with power suppressed terms. Finally, subleading
terms which are induced by SCET resummation are suppressed by powers
or logs of the hadronic scale: this feature of SCET resummation may
also be a limitation, because the partonic and hadronic scales, though
related, do not coincide, and in fact it may well be that the former
is close to threshold while the latter is not~\cite{Bonvini:2010tp}.

\section{Threshold resummation at fixed logarithmic accuracy}
\label{sec:thres}

For definiteness, we will concentrate on the production of Drell-Yan pairs
at hadron colliders. This choice does not entail loss of generality,
and the extension to other processes is straightforward.
We will consider in particular the invariant mass
distribution 
$\frac{d\sigma_{\rm DY}}{dM^2}$, with $M$ 
the invariant mass of the pair.
We define the hadronic scaling variable
\beq
\tau=\frac{M^2}{s}
\label{hadscale}
\eeq
where $s$ is the hadronic center-of-mass energy squared, so the
threshold limit is $\tau\to1$.
 Perturbative QCD factorization takes
the form
\beq
\sigma(\tau,M^2)=\int_\tau^1\frac{dz}{z}\,C(z,M^2)
{\cal L}\(\frac{\tau}{z}\),
\label{sigma}
\eeq
where ${\cal L}$ is the parton luminosity, and $\sigma(\tau,M^2)$ is a
dimensionless cross-section
\beq
\sigma(\tau,M^2)=\frac{1}{\tau \sigma_0} \frac{d\sigma_{\rm DY}}{dM^2}
\label{dyxsect}
\eeq
defined by requiring that at the Born
level (i.e.\  at order $\alpha_s^0$) $C(z,M^2)=\delta(1-z)$.
Note that Eq.~(\ref{sigma}) is a schematic expression:
in general, a sum over different parton subprocesses must be included.
In the sequel, without significant loss of generality,  
we shall always choose the renormalization and
factorization scales equal to each other and to the physical hard
scale $\mu^2_F=\mu^2_R=M^2$. 

\subsection{Perturbative QCD: resummation in $N$ space}

As discussed in Section~\ref{sec:intro}, standard QCD resummation is
more conveniently performed by taking a Mellin transform
\beq
\sigma(N,M^2)=\int_0^1d\tau\,\tau^{N-1}\sigma(\tau,M^2)
=C(N,M^2){\cal L}(N)
\label{melsigma}
\eeq
which factorizes both the convolution Eq.~(\ref{sigma}) and the gluon
radiation phase space. In Eq.~(\ref{melsigma}) by slight abuse of notation
we denote  with  $C(N,M^2)$ and ${\cal L}(N)$ the Mellin transforms of
$C(z,M^2)$ and ${\cal L}(z)$ respectively.

The $N$-space resummed coefficient function has the form
\beq
\Cqcd(N,M^2)=\bar g_0(\as)\exp\bar\S\left(M^2,\frac{M^2}{N^2}\right)
\label{Ctrad}
\eeq
where
\beq
\bar\S\left(M^2,\frac{M^2}{N^2}\right)
=\int_0^1dz\,z^{N-1}
\[\frac{1}{1-z}
\int_{M^2}^{M^2(1-z)^2} \frac{d\mu^2}{\mu^2} 2A\(\as(\mu^2)\)
+ D\(\as([1-z]^2M^2)\) \]_+.
\label{sqcddef}
\eeq
The functions $\bar g_0(\as)$, $A(\as)$ and $D(\as)$ are given as
power series in $\as$, with $\bar g_0(0)=1$ and $A(0)=D(0)=0$;
$A(\as)$ is order by order the coefficient of the soft singularity in
the Altarelli-Parisi splitting function for the relevant partonic
subprocess, while the functions $D(\as)$ and $\bar g_0(\as)$ are
process-dependent.  Specifically, in the case of Drell-Yan production
initiated by quark-antiquark collisions, the relevant Altarelli-Parisi
splitting function is
\beq
P_{qq}(\alpha_s,x)=\frac{A(\as)}{(1-x)_+}\left[1+O(1-x)\right].
\eeq

As a result, the resummed coefficient function takes the form (using
the notation of Ref.~\cite{Catani:1996yz})
\begin{align}
\Cqcd(N,M^2) 
&= g_0(\as) \exp \S\(\ab L, \ab\), \label{eq:Cres}\\
\S(\ab L, \ab) &= \frac{1}{\ab} g_1(\ab L) + g_2(\ab L) +
\ab g_3(\ab L) + \ab^2  g_4(\ab L) 
+ \ldots,\label{eq:S}\\
\ab&\equiv 2\as(M^2)\beta_0,\quad L\equiv\ln\frac{1}{{N}},\label{eq:ab_def}
\end{align}
where $\beta_0$ is the first coefficient of the QCD $\beta$ function,
defined as
\beq
\mu^2\frac{d\as(\mu^2)}{d\mu^2}=-\beta_0\as^2(\mu^2)+\Ord(\as^2);\qquad
\beta_0=\frac{11 C_A-2n_f}{12\pi}
\label{beta0}
\eeq
and the functions $g_i$ are of order $g_1(\ab L)=\Ord(\as^2)$ and 
$g_i(\ab L)= \Ord(\as)$ for $i>1$, and   are straightforwardly 
obtained performing the integrals in Eq.~(\ref{sqcddef}), and thus
each determined by a finite number of coefficients in the expansion of
the functions $A$ and $D$. Note that the functions $g_0$ and $\S$ do not
coincide with $\bar g_0$ and $\bar \S$ of Eq.~(\ref{Ctrad}), because,
by definition, $\S$ unlike $\bar\S$ does not contain terms which are not
logarithmically enhanced, while $g_0$ includes non-logarithmic 
contributions both from $\bar g_0$ itself, and from 
the integral Eq.~(\ref{sqcddef}).  

The standard perturbative QCD resummation predicts correctly all
contributions to $\ln \Cqcd(N,M^2)$ up to a given order: in other
words, if contributions up to $g_n$ are included in $\S(\ab L, \ab)$
Eq.~(\ref{eq:S})  then $\ln\Cqcd(N,M^2)$ is determined up to subleading
corrections of order $\Ord(\as^{k+(n-1)} L^k)$. This is standardly called
N$^{n-1}$LL resummation. However,  once the exponential is expanded
out in order to obtain the coefficient function $\Cqcd(N,M^2)$, at
each order in $\as$, only a restricted number of logarithmically
enhanced terms is predicted correctly, and furthermore, 
inclusion of the prefactor $g_0(\as)$ Eq.~(\ref{eq:Cres}) 
(which is not logarithmically
enhanced) is mandatory in order to improve the accuracy beyond NLL.
In fact its inclusion already at the NLL level increases the
number of contributions to the coefficient function which are
predicted correctly. In Tab.~\ref{tab:QCDcount} we summarize 
the order up to
which the expansion of the functions Eq.~(\ref{eq:Cres},\ref{eq:S})
should be included to achieve a given logarithmic accuracy, 
and, in the last column, the order of the
contributions to the resummed coefficient function $\Cqcd(N,M^2)$
which, as a consequence, are predicted correctly.
\begin{table}[htb]
\centering
\begin{tabular}[c]{c c c r}
log approx. & $g_i$ up to & $g_0$ up to order & accuracy: $\as^nL^k$\\
\midrule
    LL & $i=1$ & $(\as)^0$ & $k=2n$ \\
    NLL & $i=2$ & $(\as)^1$ & $2n-2\leq k\leq 2n$ \\
    NNLL & $i=3$ & $(\as)^2$ & $2n-4\leq k\leq 2n$\\
\midrule
\end{tabular}
\caption{Orders of logarithmic approximations and 
accuracy of the predicted logarithms in perturbative QCD.}
\label{tab:QCDcount}
\end{table}

\subsection{The SCET approach}

Resummation in SCET in the approach of Ref.~\cite{Becher:2006nr},
which henceforth we will refer to as SCET resummation  for short, 
is directly given in the physical space of momentum
fractions. The relevant expression for Drell-Yan pair production 
has been computed in Ref.~\cite{Becher:2007ty}, and it is given by
\beq
\Cscet(z,M^2,\mu_s^2) = H(M^2)U(M^2, \mu_s^2) S(z, M^2,\mu_s^2)
\label{Cscet}
\eeq
where $H(M^2)$, the so-called hard function, has an expansion in powers
of $\as$ computed at the hard scale $M^2$;
\beq
S(z,M^2, \mu_s^2) = 
\tilde s_{\rm DY}\(\ln\frac{M^2}{\mu_s^2}+\frac{\partial}{\partial\eta},\mu_s\)
  \frac{1}{1-z} \(\frac{1-z}{\sqrt{z}}\)^{2\eta} 
\frac{e^{-2\gamma \eta}}{\Gamma(2\eta)},
\eeq
where
\beq
\eta=\int_{M^2}^{\mu_s^2}\frac{d\mu^2}{\mu^2}\,\Gamma_{\rm cusp}\(\as(\mu^2)\);
\qquad \Gamma_{\rm cusp}(\as)=A(\as)
\label{etadef}
\eeq
and $\tilde s_{\rm DY}(L,\mu)$ has a perturbative expansion in powers
of $\as(\mu^2)$. Note that the function $\Gamma_{\rm cusp}(\as)$
coincides with the function
$A(\as)$ of Eq.~(\ref{sqcddef}). Finally,
\beq
U(M^2,\mu_s^2) = \exp\left\{ -\int_{M^2}^{\mu_s^2} \frac{d\mu^2}{\mu^2}
    \[\Gamma_{\rm cusp}\(\as(\mu^2)\) \ln\frac{\mu^2}{M^2}
    - \gamma_W\(\as(\mu^2)\) \]
  \right\}
\label{udef}
\eeq
where $\gamma_W(\as)$ has a power expansion in $\as$.
The resummed expression as given in Ref.~\cite{Becher:2007ty}
actually depends on several energy scales,
which here for simplicity are all taken to be equal to the hard scale $M^2$.

Two important formal aspects characterize the SCET resummed result.
The first is that it depends on a ``soft scale'' $\mu_s$, and in fact
the logs which are being resummed in SCET are
$\ln\frac{\mu_s}{M}$. Hence, different choices of soft scale lead to
different forms of the SCET resummation, as we shall discuss in greater
detail in the next Section. 

The second is related to the
well-known fact that at the endpoint $z=1$ 
the coefficient function $\Cscet(z,M^2,\mu_s^2)$ is a distribution,
rather than an ordinary function. This distribution is usually
expressed in terms of the so-called plus distribution
$\frac{1}{(1-z)}_+$. The distributional nature of the SCET result
emerges in the following way.
The convolution product of $\Cscet(z,M^2,\mu_s^2)$ with any
well-behaved test function of $z$ is
well defined as long as $\eta$ is a fixed, positive number:
the factor $(1-z)^{2\eta}$ acts as
a regulator of the soft singularity at $z=1$. The result can then be
analytically continued to negative values of $\eta$ (which is typically the case
in DY-like processes) by means of the identity
\beq
\int_0^1dz\,(1-z)^{2\eta-1}f(z)=
\int_0^1dz\,(1-z)^{2\eta-1}[f(z)-f(1)]
+\frac{1}{2\eta}f(1).
\label{reg}
\eeq
Eq.~\eqref{reg} defines a distribution on
a space of test functions $f(z)$, regular in the range $0\leq z\leq 1$,
which is usually written
\beq
(1-z)^{2\eta-1}=\[(1-z)^{2\eta-1}\]_+ +\frac{1}{2\eta}\delta(1-z).
\label{regp}
\eeq
It is important to note that $\eta$ is of order $\as$: therefore, the
term proportional to $\delta(1-z)$ in Eq.~\eqref{regp} combines with
the factor $1/\Gamma(2\eta)=2\eta+\Ord(\eta^2)$ in Eq.~\eqref{Cscet}
to form an order-$\as^0$ contribution (with the correct kinematic
structure).

As in the perturbative case, a given logarithmic accuracy is obtained
by including a finite number of terms in the perturbative expansion of
the functions which determine the resummed result, namely
$\Gamma_{\rm cusp}$, $\gamma_W$, $H$ and $\tilde s_{\rm DY}$.  The accuracy
which, according to Ref.~\cite{Becher:2007ty}, is obtained by
including in the SCET expression Eq.~(\ref{Cscet}-\ref{udef})
coefficients up to a given order, as well as the corresponding
nomenclature, are summarized in Tab.~\ref{tab:SCETcount}. As
in the case of Tab.~\ref{tab:QCDcount}, the last column gives the
order of the contributions to $\Cscet$ which are predicted exactly.
As mentioned in Section~\ref{sec:intro} and as is apparent comparing
Tab.~\ref{tab:QCDcount} to Tab.~\ref{tab:SCETcount}, beyond LL the
SCET results are always less accurate than the QCD results of the same
name: the QCD NLL includes terms of order $\alpha_s^n\ln^k\frac{\mu_s}{M}$
with $k\ge 2n-2$, but the SCET NLL only includes
terms with $k\ge 2n-1$. This was already observed in
Ref.~\cite{Ahrens:2011mw}. 

\begin{table}[htb]
\begin{center}
\begin{tabular}{ccrccc}
RG-impr.\ PT & log.\ approx.\ 
 & $\Gamma_{\rm cusp}$
 & $\gamma_W$ & $H$, $\tilde s_{\rm DY}$ & accuracy: $\alpha_s^n \ln^k\frac{\mu_s}{M}$\\
 \hline\\[-0.4cm]
 --- & LL & 1-loop
 & tree-level & tree-level & $k= 2n$ \\
 LO & NLL & 2-loop
 & 1-loop & tree-level & $2n-1\le k\le 2n$ \\
 NLO & NNLL  & 3-loop & 2-loop 
 & 1-loop & $2n-3\le k\le 2n$ \\
 NNLO & NNNLL  & 4-loop & 3-loop
 & 2-loop & $2n-5\le k\le 2n$\\[0.2cm]
\hline
\end{tabular}
\caption{
Different approximation schemes for the evaluation of the resummed 
cross-section formulae in the SCET approach.}
\label{tab:SCETcount}
\end{center}
\end{table}
When comparing the two different definitions of logarithmic accuracy,
Tab.~\ref{tab:QCDcount} and Tab.~\ref{tab:SCETcount}, one should
distinguish a purely terminological
issue and an issue of substance. The terminological issue is how each
given accuracy is called: this is clearly immaterial. The issue of
substance is whether at (say) NLL the SCET expression
Eq.~(\ref{Cscet}-\ref{udef})
may be upgraded to the higher accuracy of the NLL QCD expression 
Eq.~(\ref{Ctrad}-\ref{sqcddef}) (without having to resort to the yet
more accurate NNLL SCET
expression), and likewise at all subsequent logarithmic orders. We
will show that the answer to this question depends on the choice of
soft scale $\mu_s$.\footnote{In other contexts, such as for
  example the resummation of jet veto logs~\cite{Berger:2010xi},
  SCET results which correspond either of  two different accuracies,
  respectively akin to  Tab.~\ref{tab:SCETcount} or 
  Tab.~\ref{tab:QCDcount}, may be achieved by suitable choices of terms to
  be included in the resummed expression.}

\section{Choice of the soft scale and SCET-QCD comparison}
\label{sec:soft}

In the standard perturbative QCD approach to soft resummation, the
energy scale which characterizes soft gluon emission is of the order
of $M(1-z)$: when the observed final state carries away a fraction $z$
of the available partonic energy, the energy available for unobserved
radiation is $M(1-z)$, which is much smaller than $M$ if $z$ is close
to 1. The fact that the scale involved is partonic has
phenomenological implications: because the partonic center-of-mass
energy is always smaller than the hadronic one, threshold resummation may be
relevant even for processes which are relatively far from hadronic
threshold, provided the parton luminosity is peaked for small values
of the momentum fraction~\cite{Bonvini:2010tp}. This
indeeds is known to happen for  Higgs production in gluon fusion at
the LHC.~\cite{Catani:2001cr,Catani:2003zt}

In SCET resummation, however, one resums logs of the large 
ratio $M/\mu_s$ of the hard scale $M$ to the soft scale $\mu_s$, and
various choices for the soft scale $\mu_s$ are possible: in
particular, the choice which has been advocated in
Refs.~\cite{Becher:2006nr,Becher:2006mr,Becher:2007ty,Ahrens:2008qu},
and which removes the problem of the Landau pole, consists of choosing
for $\mu_s$ a  scale which characterizes the (hadronic) physical process.

If $\mu_s$ is chosen as a function of the partonic scaling variable
$z$, then  the resummed SCET partonic cross-section  
$\Cscet(z,M^2,\mu^2_s)$ Eq.~(\ref{Cscet})
can be directly compared to the momentum-space perturbative QCD
expression, which may be obtained by determining the inverse Mellin
transform $\Cqcd(z,M^2)$ of the resummed $N$-space expression
Eq.~(\ref{Ctrad}). We will study this case in detail in the next Section.
However, if  $\mu_s$ is chosen as a function of the hadronic scaling variable
$\tau$, the SCET and perturbative QCD resummed
results must be compared at the level of physical cross-sections
$\sigqcd(\tau,M^2)$ and $\sigscet(\tau,M^2)$, which
are respectively obtained substituting $\Cqcd(z,M^2)$  or 
$\Cscet(z,M^2,\mu^2_s)$ in the factorized expression
Eq.~(\ref{sigma}), with some particular choice of soft scale $\mu_s$.

It is important to understand that these different choices of soft
scale lead to resummed predictions with different analytic structure. 
To see this,
note that if the soft scale only depends on the parton momentum
fraction $z$, then Eq.~(\ref{sigma}) is a convolution, in the sense
that upon Mellin transformation it factorizes according to
Eq.~(\ref{melsigma}). This factorization is of course a necessary and
sufficient condition for parton radiation to respect longitudinal
momentum conservation.  But if in Eq.~(\ref{sigma}) the coefficient
function depends on $\tau$ through the soft scale, then the
convolution structure is destroyed. This means that with this
particular choice of soft scale,  upon Mellin transformation the
cross-section no longer factorizes,  thereby violating longitudinal
momentum conservation.  This also violates
the QCD factorization theorem, because the short-distance partonic cross section
depends on long-distance physics through the hadronic variable
$\tau$. The possibility of making this choice of soft scale, and
indeed the very possibility of making alternative choices of soft
scale, some of which preserve factorization and some of which do not,
appears puzzling in a standard perturbative QCD approach. We will not
attempt to address the issue of principle of understanding this
apparent structural discrepancy between SCET and QCD results. Rather,
we will take the SCET and QCD expressions at face value: our aim will
be to determine how they are related to each other.

We will now derive a master formula which relates the SCET resummed
expression for generic choice of the soft scale to the standard
perturbative QCD expression. For definiteness, we specialize to the
next-to-next-to-leading log case, but all relevant structures are
already present at this order so generalization to higher logarithmic
orders is straightforward. First, we give the explicit expression of
the QCD result Eq.~(\ref{Ctrad}) to this order. Then, we give the SCET
expression Eq.~(\ref{Cscet}-\ref{udef}) to the same order, and we
perform its (exact) Mellin transform in order to allow for a
comparison with the QCD expression, which is given in $N$
space. Finally, by comparing the two expressions we derive a master
formula which relates them, as a function of the soft scale $\mu_s$,
through a suitable factor (in Mellin space) or a convolutive function
(in momentum space).

\subsection{Perturbative QCD resummation to NNLL}
\label{sec:qcdnnll}

The NNLL resummed expression in perturbative QCD is given by
Eq.~(\ref{Ctrad}) with~\cite{Moch:2005ky,Moch:2005ba} (see also
Ref.~\cite{Bonvini:2010tp})
 \begin{align}
&A(\as)=\frac{A_1}{4}\as+\frac{A_2}{16}\as^2+\frac{A_3}{64}\as^3+\Ord(\as^4),
\label{adef}
\\
&\quad A_1=\frac{4C_F}{\pi};
\label{aone}
\\
&D(\as)=D_1\as+D_2\as^2+\Ord(\as^3),
\label{ddef}
\\
&\quad D_1=0,\quad
D_2 = \frac{C_F}{16\pi^2}
\left[ C_A\(-\frac{1616}{27}+\frac{88}{9}\pi^2+56\zeta_3\) 
+ \(\frac{224}{27}-\frac{16}{9}\pi^2\) n_f \right].
\label{donetwo}
\end{align}
We can perform the $z$ integral in Eq.~(\ref{Ctrad})
using Eq.~(\ref{MellinNNLL2}):
\begin{align}
\bar{\cal S}_{\rm\scriptscriptstyle QCD}\(M^2,\frac{M^2}{\bar{N}^2}\)
&=\int_{M^2}^{M^2/\bar{N}^2}\frac{d\mu^2}{\mu^2}\, \[A\(\as(\mu^2)\) 
\(\ln\frac1{\bar{N}^2} - \ln\frac{\mu^2}{M^2}\)+\frac{1}{2}D\(\as(\mu^2)\)\]
\nonumber\\
&\qquad+\frac{\pi^2}{12}\frac{d^2}{dL^2}
\int_{M^2}^{M^2/N^2}\frac{d\mu^2}{\mu^2}\, \[A\(\as(\mu^2)\) 
\(\ln\frac1{N^2} - \ln\frac{\mu^2}{M^2}\)+\frac{1}{2}D\(\as(\mu^2)\)\]
\nonumber\\
&=\int_{M^2}^{M^2/\bar{N}^2}\frac{d\mu^2}{\mu^2}\, \[A\(\as(\mu^2)\) 
\(\ln\frac1{\bar{N}^2} - \ln\frac{\mu^2}{M^2}\)+\frac{1}{2}D\(\as(\mu^2)\)\]
\nonumber \\
&\qquad+\frac{C_F\pi}{3}\as\(\frac{M^2}{\bar{N}^2}\),
\label{sexpra}
\end{align}
where (as per Eq.~(\ref{Lbar})) $\bar N=Ne^\gamma$.
We have
neglected subleading (N$^3$LL) terms (including the
replacement $N\to\bar{N}$ in the argument of $\as$
in the last term) and we have brought all integrals
to a common form using
\begin{align}
\int_0^{1-1/N}\frac{dz}{1-z} 
2\int_{M^2}^{M^2(1-z)^2} \frac{d\mu^2}{\mu^2}\, A\(\as(\mu^2)\)
&=-\int_{M^2}^{M^2/N^2}\frac{d\nu^2}{\nu^2} 
\int_{M^2}^{\nu^2} \frac{d\mu^2}{\mu^2}\, A\(\as(\mu^2)\)\nonumber \\
&=-\int_{M^2}^{M^2/N^2}\frac{d\mu^2}{\mu^2}\, A\(\as(\mu^2)\) 
\(\ln\frac1{N^2} - \ln\frac{\mu^2}{M^2}\)
\label{atermint}
\\
\int_0^{1-1/N}\frac{dz}{1-z}\,D\(\as(M^2(1-z)^2\)
&=-\frac{1}{2} \int_{M^2}^{M^2/N^2}\frac{d\mu^2}{\mu^2}\,
D\(\as(\mu^2)\).
\label{dtermint}
\end{align}

In order to ease the subsequent comparison to the SCET result,  we separate off the
non-logarithmic constant from the last term in Eq.~(\ref{sexpra}):
\begin{align}
\bar{\cal S}_{\rm\scriptscriptstyle QCD}\(M^2,\frac{M^2}{\bar{N}^2}\)
&=\int_{M^2}^{M^2/\bar{N}^2}\frac{d\mu^2}{\mu^2}\,
\[A\(\as(\mu^2)\)\(\ln\frac1{\bar{N}^2}-\ln\frac{\mu^2}{M^2}\)
+\hat D_2\as^2(\mu^2)\]
\nonumber\\
&\qquad +\frac{C_F\pi}{3}\as(M^2)
\label{sexprb}
\end{align}
where
\beq
\hat D_2=\frac{D_2}{2}-\frac{C_F\pi}{3}\beta_0
=\frac{C_F}{16\pi^2}\[ C_A\(-\frac{808}{27}+28\zeta_3\)
+\frac{112}{27}n_f\].
\label{dhatdef}
\eeq
We can thus write 
\beq
\Cqcd(N,M^2)=\hat g_0(\as(M^2))\exp\hat{\cal S}_{\rm\scriptscriptstyle QCD}
\(M^2,\frac{M^2}{\bar{N}^2}\),
\label{CQCD}
\eeq
where
\begin{align}
&\hat g_0(\as)=1+\hat g_{01}\as+\Ord(\as^2);
\label{ghatdef}\\
&\hat{\cal S}_{\rm\scriptscriptstyle QCD}\(M^2,\frac{M^2}{\bar{N}^2}\)
=\int_{M^2}^{M^2/\bar{N}^2}\frac{d\mu^2}{\mu^2}\,
\[A\(\as(\mu^2)\)\(\ln\frac1{\bar{N}^2}-\ln\frac{\mu^2}{M^2}\)
+\hat D_2\as^2(\mu^2)\].
\label{shatdef}
\end{align}
Note that $\hat g_0$ and $\hat{\cal S}$ cannot be identified with $g_0$
and $\S$ in Eq.~(\ref{Ctrad}), because the integral in
Eq.~(\ref{shatdef}) does contain some terms which are not
logarithmically enhanced:
\beq
\frac{A_1}{4}\int_{M^2}^{M^2/\bar{N}^2}\frac{d\mu^2}{\mu^2}\,\as(\mu^2)
\(\ln\frac{1}{\bar{N}^2}-\ln\frac{\mu^2}{M^2}\)
=\frac{2C_F}{\pi}\gamma^2\as(M^2)+\text{logarithms}+\Ord(\as^2),
\label{constint}
\eeq
so that 
\beq
\hat g_{01}=g_{01}-\frac{2C_F}{\pi}\gamma^2.
\label{hatgzone}
\eeq
However, the form
Eq.~\eqref{shatdef} of the exponent in the QCD result
is especially suited
for comparison to the SCET result, as we now show.

\subsection{SCET resummation to NNLL}
\label{sec:scetnnll}

We turn to the SCET expression, which is given by Eq.~(\ref{Cscet})
with, to NNLL
\begin{align}
&\gamma_W(\as)=\gamma_W^{(2)}\frac{\as^2}{16\pi^2}+\Ord(\as^3),
\label{gamwdef}\\
&\quad \gamma_W^{(2)}=C_F C_A\(-\frac{808}{27}+\frac{11\pi^2}{9}+28\zeta_3\)
+C_F T_F n_f\(\frac{224}{27}-\frac{4\pi^2}{9}\).
\label{gamwtwo}
\end{align}
In order to compare it to the perturbative QCD result, we perform a
Mellin transform with respect to $z$. This is easy to do, because
the $z$ dependence is all contained in the soft function
$S(z,M^2,\mu_s^2)$, whose Mellin transform is
\begin{align}
{\cal M}\[S(z,M^2,\mu_s^2)\] &= 
\tilde s_{\rm DY}\(\ln\frac{M^2}{\mu_s^2}+\frac{\partial}{\partial\eta},\mu_s\)
\frac{\Gamma(N-\eta)\Gamma(2\eta)}{\Gamma(N+\eta)}
\frac{e^{-2\gamma \eta}}{\Gamma(2\eta)}
\nonumber\\
&=\[1+\frac{C_F}{2\pi}\as(\mu_s^2)
\(\ln^2\frac{M^2}{\mu_s^2\bar{N}^2}
+\frac{\pi^2}{6}\)\]\bar{N}^{-2\eta}+\Ord\(\frac{1}{N}\).
\label{smellin}\end{align}
It follows that the Mellin transform of the coefficient function
Eq.~(\ref{Cscet}) is
\begin{align}
&\Cscet(N,M^2,\mu_s^2) = H(M^2)
\[1
+\frac{C_F\pi}{12}\as(\mu_s^2)
+\frac{C_F}{2\pi}\as(\mu_s^2)
\ln^2\frac{M^2}{\mu_s^2\bar{N}^2}
\]
\nonumber\\
&\qquad\times\exp\int_{M^2}^{\mu_s^2} \frac{d\mu^2}{\mu^2}
\[\Gamma_{\rm cusp}\(\as(\mu^2)\)
\(\ln\frac{1}{\bar{N}^2}- \ln\frac{\mu^2}{M^2}\)
    +\frac{\gamma_W^{(2)}}{16\pi^2}\as^2(\mu^2) \]
+\Ord\(\frac{1}{N}\).
\label{eq:SCET_Nnll}
\end{align}

It is very important to observe that the Mellin transform has been
computed {\it at fixed $\mu_s$}. This means that firstly,
Eq.~(\ref{eq:SCET_Nnll}) is  not the Mellin transform of the SCET
expression when $\mu_s$ depends on $z$ (which we  will discuss in the next
Section): in that case the Mellin transform would also act on the $z$
dependence through $\mu_s$. And second, that if
$\mu_s$ depends on $\tau$ the cross-section $\sigscet(\tau,
M^2)$ 
computed using Eq.~(\ref{sigma}) does not factorize into the
product of $\Cscet(N,M^2,\mu_s^2)$ Eq.~(\ref{eq:SCET_Nnll})
times a parton luminosity ${\cal L}(N)$  upon Mellin transformation:
the Mellin integral over $\tau$ would also act on the $\tau$
dependence through $\mu_s$ which, as already noted, does not have the
form of a convolution integral.

Equation~(\ref{eq:SCET_Nnll}) can be brought in a form which is suitable
for comparison to the QCD expression  by separating off the constant
as in Eq.~(\ref{sexprb}), thus leading to
\beq
\Cscet(N,M^2,\mu_s^2)=\hat H(M^2)E(N,M^2,\mu_s^2)
\exp\hat{\cal S}_{\rm\scriptscriptstyle SCET}(M^2,\mu_s^2),
\label{Cscetnew}
\eeq
with
\begin{align}
&\hat H(M^2)=H(M^2)\exp\[\frac{C_F\pi}{12}\as(M^2)\]
=1+\as(M^2)\(H_1+\frac{C_F\pi}{12}\)+\Ord(\as^2);
\label{hdef}\\
&\hat{\cal S}_{\rm\scriptscriptstyle SCET}(M^2,\mu_s^2)=
\int_{M^2}^{\mu_s^2} \frac{d\mu^2}{\mu^2}
\[\Gamma_{\rm cusp}\(\as(\mu^2)\)
\(\ln\frac{1}{\bar{N}^2}- \ln\frac{\mu^2}{M^2}\)
+\hat\gamma_W^{(2)}\as^2(\mu^2) \],
\label{scetsfun}\\
&\qquad \hat\gamma_W^{(2)}=\frac{\gamma_W^{(2)}}{16\pi^2}-\frac{C_F\pi}{12}\beta_0
=\frac{C_F}{16\pi^2}\[ C_A\(-\frac{808}{27}+28\zeta_3\)
+\frac{112}{27}n_f\];
\label{gamtwohat}\\
&E(N,M^2,\mu_s^2)=1+\frac{C_F}{2\pi}\as(\mu_s^2)
\(\ln\frac{1}{\bar{N}^2}-\ln\frac{\mu_s^2}{M^2}\)^2.
\label{efun}\end{align}

\subsection{The master formula}
\label{sec:master}
The QCD expression Eqs.~(\ref{CQCD}-\ref{shatdef}) and the SCET
expression Eqs.~(\ref{Cscetnew}-\ref{efun}) are easily related, by
noting that, because  $\Gamma_{\rm cusp}(\as)=A(\as)$ and
$\hat D_2=\hat\gamma_W^{(2)}$, the integrands in Eqs.~(\ref{shatdef})
and~(\ref{scetsfun}) coincide, so
\beq
\hat{\cal S}_{\rm\scriptscriptstyle SCET}(M^2,Q^2)
=\hat{\cal S}_{\rm\scriptscriptstyle QCD}(M^2,Q^2)
\equiv\hat{\cal S}(M^2,Q^2).
\label{sexp}
\eeq
It follows that, splitting the integral as
 $\int_{M^2}^{M^2/\bar{N}^2}\frac{d\mu^2}{\mu^2}=\int_{M^2}^{\mu_s^2}\frac{d\mu^2}{\mu^2}+\int_{\mu_s^2}^{M^2/\bar{N}^2}\frac{d\mu^2}{\mu^2}$,
we get
\beq
\Cqcd(N,M^2)=C_r(N,M^2,\mu^2_s)\Cscet(N,M^2,\mu_s^2)
\label{CCC}
\eeq
where
\beq
C_r(N,M^2,\mu_s^2)=\frac{\hat g_0(\as(M^2))}{\hat H(M^2)}
\frac{\exp\hat{\cal S}\(\mu_s^2,\frac{M^2}{\bar{N}^2}\)}{E(N,M^2,\mu_s^2)}.
\label{Cr}
\eeq

The non-logarithmic terms in fact cancel to the accuracy of our
computation. 
Indeed, 
by substituting the value~\cite{Moch:2005ba}
\beq
g_{01}=\frac{C_F}{\pi}\left( 4\zeta_2 -4 + 2\gamma^2\right) 
\label{goneval}
\eeq
in Eq.~(\ref{hatgzone}), and the value~\cite{Becher:2007ty}
\beq
H_1=\frac{C_F}{\pi}\(\frac{7}{2}\zeta_2-4\)
\label{honeval}
\eeq
in Eq.~\eqref{hdef} we get 
\beq
\frac{\hat g_{0}(\as(M^2))}{\hat H(M^2)}=1+\Ord(\as^2),
\label{constsimp}
\eeq
so deviations from unity are of the same order as the first
contribution which, at NNLL accuracy, 
is not included in $H(M^2)$ (according to
Tab.~\ref{tab:SCETcount}). 
The expression of $C_r$ can be further simplified by including the
function $E(N,M^2,\mu_s^2)$ Eq.~(\ref{efun}) in the function $\hat\S$:
indeed
\begin{align}
&E(N,M^2,\mu_s^2)=\exp\[\frac{A_1}{8}\as(\mu_s^2)
\(\ln\frac1{\bar{N}^2}-\ln\frac{\mu_s^2}{M^2}\)^2\]+\Ord(\as^2)
\nonumber\\
&=\exp\int_{\mu_s^2}^{M^2/\bar{N}^2}\frac{d\mu^2}{\mu^2}\,\Bigg[
\frac{A_1}{4}\as(\mu^2)\(\ln\frac1{\bar{N}^2}-\ln\frac{\mu^2}{M^2}\)
-\frac{A_1}{8}
\beta(\as(\mu^2))
\(\ln\frac1{\bar{N}^2}-\ln\frac{\mu^2}{M^2}\)^2\Bigg].
\label{Eexp}
\end{align}

Using Eq.~\eqref{Eexp} in the definition of
$C_r(N,M^2,\mu_s^2)$ we obtain our final expression 
\beq
C_r(N,M^2,\mu_s^2)=\exp\hat{\cal S}_r
\(\mu_s^2,\frac{M^2}{\bar{N}^2}\),
\label{crfin}
\eeq
with
\begin{align}
\hat{\cal S}_r\(\mu_s^2,\frac{M^2}{\bar{N}^2}\)
&=\int_{\mu_s^2}^{M^2/\bar{N}^2}\frac{d\mu^2}{\mu^2}\,
\Bigg[\(A(\as(\mu^2))-\frac{A_1\as(\mu^2)}{4}\)
\(\ln\frac1{\bar{N}^2}-\ln\frac{\mu^2}{M^2}\)
\nonumber\\
&\qquad+\frac{A_1}{8}
\beta(\as(\mu^2))
\(\ln\frac1{\bar{N}^2}-\ln\frac{\mu^2}{M^2}\)^2+\hat D_2\as^2(\mu^2)
\Bigg].
\label{Sr}
\end{align}
Equation~(\ref{CCC}) together with the explicit expression
Eqs.~(\ref{crfin}-\ref{Sr}) of the function $C_r$ provides the master
formula which relates SCET and perturbative QCD resummation. It is the
main result of this paper. 
We note that no term of order $\as$ appears in the integrand
of Eq.~\eqref{Sr}: indeed, the inclusion of  the term $E(N,M^2,\mu_s^2)$ has
the effect of removing the term proportional to $A_1\alpha_s$ 
(see Eq.~\eqref{adef}). The
remaining contributions to the integrand in Eq.~(\ref{Sr}) start at
$\Ord(\as^2)$.

It is important to observe that while  $\Cqcd(N,M^2)$
does not admit
a Mellin inverse,
because it has a cut in the complex $N$ plane starting at the value
$N_L$ at which the strong coupling blows up, $\Cscet(N,M^2,\mu_s^2)$
does admit a Mellin inverse as long as $\mu_s$
is kept fixed, because the argument of the strong coupling in the SCET
expression does not depend on $N$. This means that if Eq.~(\ref{CCC}) 
is expanded in powers of $\alpha_s(M^2)$, and then the expansion is
Mellin-inverted term by term,  the expansion of the left-hand side is
divergent, while on the right-side the Mellin inverse
of the expansion of 
$\Cscet(N,M^2,\mu_s^2)$ converges to 
$\Cscet(z,M^2,\mu_s^2)$  Eq.~(\ref{Cscet}). Therefore, the
divergence has been isolated in the Mellin inverse of the expansion of
the function $C_r(N,M^2,\mu_s^2)$ Eq.~(\ref{Cr}).

If the perturbative expansion of both sides of Eq.~(\ref{CCC}) in
powers of $\alpha_s(M^2)$ is truncated to any finite order, then the
Mellin inverse of both sides exists, and one gets the momentum-space
relation
\beq
\Cqcd(z,M^2)=\int_z^1\frac{dy}{y}C_r\left(\frac{y}{z},M^2,\mu_s^2\right)
\Cscet\left(y,M^2,\mu_s^2\right),
\label{CCCmom}
\eeq
where $\Cscet(z,M^2,\mu_s^2)$  is given by Eq.~(\ref{Cscet}) 
(expanded out to the given order), while both
$\Cqcd(z,M^2)$ and $C_r\left(z,M^2,\mu_s^2\right)$ should be
understood as the truncation to the given order of the Mellin inverse
of the expansion of the corresponding $N$--space quantities.
Equation~(\ref{CCCmom}) is then the momentum-space version of the
master formula.  

The master formula Eqs.~(\ref{CCC},\ref{CCCmom})  has been established at
next-to-next-to-leading logarithmic order, defined according to
Tab.~\ref{tab:SCETcount}. Note, however,  that the accuracy is
upgraded to 
the higher one  of Tab.~\ref{tab:QCDcount} if non-logarithmic
terms cancel to $\Ord(\as^3)$, i.e.\  if the function $\hat H(M^2)$ in
the SCET coefficient function Eq.~(\ref{Cscetnew}) is replaced by a
function $\bar H(M^2)$ such that
\beq
\frac{\hat g_{0}(\as(M^2))}{\bar H(M^2)}=1+\Ord(\as^3).
\label{constsimpcub}
\eeq
Of course, this can always be achieved by 
letting $\bar H(M^2)= \hat H
(M^2)+ \bar H_2\as^2(M^2)$ and suitably choosing the value of $\bar
H_2$, while including the $\Ord(\as^2)$ to $g_{0}(\as(M^2))$, as per
Tab.~\ref{tab:QCDcount}.   
(Whether $\bar H(M^2)$ coincides with the  $\Ord(\as^2)$
expression of $\hat H(M^2)$ as obtained using SCET 
is an issue that we will not address here).
We conclude that the master formula holds up to NNLL accuracy, defined
as in Tab.~\ref{tab:QCDcount}. It is easy to convince oneself that
this argument should hold to all logarithmic orders.

\section{Perturbative QCD vs. SCET: partonic cross-sections}
\label{sec:partonic}

The master formula Eqs.~(\ref{CCC}-\ref{CCCmom}) shows how
SCET resummation can be used to reproduce standard
results. Indeed, it
immediately implies that if we fix the soft scale in terms of the
Mellin-space  variable, 
\beq
\mu_s=\frac{M}{\bar{N}},
\label{muspartonicmel}
\eeq
then $C_r(N,M^2,\mu^2_s)=1$, i.e.\ 
\beq
\Cqcd(N,M^2)=\Cscet\(N,M^2,\frac{M^2}{\bar{N}^2}\),
\label{Mellincomp}
\eeq
so the standard QCD result is reproduced:
with this choice, SCET resummation is performed at the level of
Mellin-space partonic cross-sections. Notice that because with this
choice the SCET and QCD expressions coincide, they also have the same
accuracy. So with this choice the SCET results actually has the
accuracy of Tab.~\ref{tab:QCDcount}, rather than the lower accuracy of
Tab.~\ref{tab:SCETcount}. The equivalence of Mellin-space SCET
resummation to the  QCD
expressions was already established in Ref.~\cite{Idilbi:2005ni}; it was also
pointed out in Ref.~\cite{Becher:2007ty}, but
with the lower accuracy of Tab.~\ref{tab:SCETcount}.

Alternatively, one may try to use SCET resummation for partonic
cross-sections, but using the momentum-space SCET formula
Eq.~(\ref{Cscet}), with $\mu_s$ fixed as a momentum-space partonic scale, namely
\beq
\mu_s=M(1-z).
\label{muspartonic}
\eeq
This choice for instance was adopted recently in
Ref.~\cite{Beneke:2011mq} to perform threshold resummation for top
production. This choice also provides another way of re-deriving the
standard perturbative resummation from SCET. Indeed, it can be shown
that, away from the endpoint $z=1$,
all logarithmically enhanced terms $\frac{\ln^p(1-z)}{1-z}$ in the
partonic cross-section are reproduced order by order with this choice.

This is very easily seen at the leading-log, fixed-coupling
level. Indeed, in this limit one has
\beq
\eta=\frac{\as A_1}{2}\ln(1-z),\qquad
(1-z)^{2\eta}=\exp\[\as A_1\ln^2(1-z)\],
\label{etaafix}\eeq
so that 
\begin{align}
\Cscet(z,M^2,M^2(1-z)^2)&=\exp\[-\frac{A_1\as}{4}
\int_{M^2}^{M^2(1-z)^2} \frac{d\mu^2}{\mu^2}
\ln\frac{\mu^2}{M^2}\]\frac{(1-z)^{2\eta}}{1-z} 
\frac{1}{\Gamma(2\eta)}
\nonumber\\
&=\exp\[ -\frac{A_1\as}{2}\ln^2(1-z)\]\frac{(1-z)^{2\eta}}{1-z}
\frac{1}{\Gamma(2\eta)}.
\label{cscetz}
\end{align}
But to leading log order one may  expand $1/\Gamma(2\eta)$ to
first order in $\as$, so 
\beq
\Cscet(z,M^2,M^2(1-z)^2)
=\as A_1\frac{\ln(1-z)}{1-z}\exp\[\frac{A_1\as}{2}\ln^2(1-z)\]
+\text{NLL};\quad z\ne 1.
\label{Cscetalphafix}
\eeq
On the other hand, the perturbative result in the same approximation
is the inverse Mellin transform of
\beq
\Cqcd(N,M^2) = \exp\[\frac{\as A_1}2 \ln^2\frac{1}{N}\]+\text{NLL},
\label{llalphafix}
\eeq
i.e., using the results of Appendix~\ref{sec:applus},
\beq
\Cqcd(z,M^2)=\frac{1}{1-z}
\left.\exp\(\frac{\as A_1}{2}\frac{\partial^2}{\partial\xi^2}\)
\frac{(1-z)^\xi}{\Gamma(\xi)}
\right|_{\xi=0}+\text{NLL};\quad z\ne 1.
\label{cct}
\eeq
Expanding the exponential and keeping only
leading log terms this is seen to coincide with
Eq.~(\ref{Cscetalphafix}).

However, as pointed out in Ref.~\cite{Catani:1996yz} and discussed in
Section~\ref{sec:intro}, $\Cqcd(z,M^2)$ (defined as the leading-log
truncation of the inverse Mellin of Eq.~(\ref{llalphafix})) is
ill-defined at the endpoint $z=1$: it behaves as a distribution which
leads to a divergent integral upon convolution with any reasonably
behaved luminosity, and, if expanded order by order in $\alpha_s$, it
diverges factorially.  The SCET expression, either in the form of
Eq.~(\ref{cscetz}) or of Eq.~\eqref{Cscetalphafix}, is also
ill-defined as $z\to1$. Indeed, because now $\eta$ depends on $z$ (see
Eq.~(\ref{etaafix})), it is no longer possible to use Eq.~(\ref{regp})
to regulate the behaviour of $\Cscet(z,M^2,\mu_s)$. Note that
Eq.~(\ref{regp}) also had the effect of generating the required
$\Ord(\as^0)$ contribution to $\Cscet(z,M^2,M^2(1-z)^2)$ proportional
to $\delta(1-z)$.  Furthermore, as $z\to1$ the coefficient function
Eq.~(\ref{cscetz}) oscillates with a factorially-growing amplitude,
because of the factor $\frac{1}{\Gamma(2\eta)}$.  The fact that the
SCET resummed expression diverges at the partonic endpoint was already
noticed in Ref.~\cite{Beneke:2011mq}. Because of these difficulties,
we will not pursue further the choice Eq.~(\ref{muspartonic}) of soft
scale.

\section{Perturbative QCD vs. SCET in momentum space:
hadronic cross-sections}
\label{sec:hadronic}

We now turn finally to the choice of soft scale which is recommended
in Refs.~\cite{Becher:2006nr,Becher:2006mr,Becher:2007ty,Ahrens:2008qu},
specifically as a solution to the problem of the Landau pole, 
namely, a soft scale fixed in terms of the hadronic momentum
scale\footnote{In
  Refs.~\cite{Becher:2006nr,Becher:2006mr,Becher:2007ty,Ahrens:2008qu}
  a slightly more general choice of soft scale is considered: namely,
the soft scale Eq.~\eqref{softhad} is generally rescaled
by a function of $\tau$ which does not vanish at $\tau=1$, and is
chosen in such a way that the finite-order perturbative 
expansion of $\tilde s_{\rm DY}$ is reliable. Because this modification
does not introduce any extra logarithmic enhancement, it does not
affect our discussion, and we will not consider it.}
\beq
\mu_s=M(1-\tau).
\label{softhad}
\eeq
With this choice of soft scale, 
the SCET and perturbative QCD results can only be
compared at the level of hadronic cross-sections
\begin{align}
&\sigqcd(\tau,M^2)=\int_\tau^1\frac{dz}{z}\,\Cqcd(z,M^2)
{\cal L}\(\frac{\tau}{z}\),
\label{sigmaQCD}
\\
&\sigscet(\tau,M^2)=\int_\tau^1\frac{dz}{z}\,
\Cscet(z,M^2,\mu_s^2){\cal L}\(\frac{\tau}{z}\).
\label{sigmascet}
\end{align}
Indeed, with the choice of soft scale Eq.~(\ref{softhad}) 
the resummed SCET cross-section Eq.~\eqref{sigmascet} is no longer in the
form of a convolution product, because the integrand depends on $\tau$
explicitly in the lower integration bound and in the
argument of ${\cal L}$, but also implicitly through $\mu_s^2$. As a
consequence, upon Mellin transformation with respect to $\tau$,
$\sigscet(\tau,M^2)$, unlike the standard QCD result, 
does not factorize into a parton
luminosity and a partonic cross-section. 

Therefore, the comparison
must be carried out directly at the level of hadronic cross-sections
Eqs.~(\ref{sigmaQCD}-\ref{sigmascet}),
using the momentum-space form Eq.~(\ref{CCCmom}) of the master
formula (always understood as a truncation to arbitrary but finite order in
$\alpha_s$, as discussed in the end of Section~\ref{sec:master}).
This is somewhat problematic, because the power counting of
Tabs.~\ref{tab:QCDcount}-\ref{tab:SCETcount} was defined at the level
of coefficient functions and thus necessarily at the level of a
partonic cross-section. Of course, it is possible to define a given
logarithmic order at the level of SCET coefficient functions, then use
this expression to compute the cross-section $\sigscet(\tau,M^2)$
using Eq.~(\ref{sigmascet}). However, because this expression is not
factorized, the question whether $\sigscet(\tau,M^2)$ and
$\sigqcd(\tau,M^2)$ agree at any given order can only be answered by
comparing them directly, and counting logs of the hadronic scale
$1-\tau$. The result will then inevitably depend on the choice of
parton distributions. The only alternative is to simply conclude that
the SCET result with this choice cannot be compared to the
perturbative one, and cannot be endowed with a perturbative
meaning~\cite{Catanipriv}.

We will perform this comparison by computing the difference between 
$\sigscet(\tau,M^2)$ and
$\sigqcd(\tau,M^2)$ up to $\Ord(\as^2(M^2))$ and using the master
formula to relate results. We will then discuss the structure of the
result to all orders.

\subsection{Fixed-order comparisons}
\label{sec:fixord}

We start by computing
the function $C_r(N,M^2,\mu_s^2)$ Eq.~(\ref{crfin}) explicitly. Up to
order $\as^2$ we find
\beq
C_r(N,M^2,\mu_s^2)=1+\as^2(M^2)
\(-\frac{A_1}{3}\beta_0\ln^3\frac{c}{N}
+\frac{A_2}{8}\ln^2\frac{c}{N}+2\hat D_2\ln\frac{c}{N}\)
+\Ord(\as^3)
\label{CrNa2}
\eeq
where
\beq
c=\frac{Me^{-\gamma}}{\mu_s}.
\label{cdef}
\eeq
The corresponding momentum-space expression is readily obtained by
performing the  inverse Mellin transform of Eq.~\eqref{CrNa2} with the
help of Eq.~\eqref{MellinlogcN}:
\begin{align}
C_r(z,M^2,\mu_s^2)&=\delta(1-z)
\nonumber\\
&\qquad+\left.
\as^2(M^2)
\(
-\frac{A_1}{3}\beta_0\frac{\partial^3}{\partial\xi^3}
+\frac{A_2}{8}\frac{\partial^2}{\partial\xi^2}
+2\hat D_2\frac{\partial}{\partial\xi}\)
c^\xi K(z,\xi)\right|_{\xi=0}+\Ord(\as^3),\label{CCCNNLL2}
\end{align}
where the function
\beq
K(z,\xi)=\Delta(\xi)\ln^{\xi-1}\frac{1}{z}
\label{Kdef}
\eeq
plays the role of a generating function.

The difference between the resummed physical cross-sections in the QCD
and SCET formalisms is now found substituting the explicit expression
of $C_r$ Eq.~\eqref{CCCNNLL2} in the master formula
Eq.~(\ref{CCCmom}):
\beq
\sigqcd(\tau,M^2)=\sigscet(\tau,M^2)
+\as^2(M^2)\left.\(
-\frac{A_1}{3}\beta_0\frac{\partial^3}{\partial\xi^3}
+\frac{A_2}{8}\frac{\partial^2}{\partial\xi^2}
+2\hat D_2\frac{\partial}{\partial\xi}
\)
c^\xi \Sigma(\tau,\xi)\right|_{\xi=0}
\label{tappeto}
\eeq
where
\begin{align}
\Sigma(\tau,\xi)
&=\int_\tau^1\frac{dz}{z}\,K(z,\xi)\sigscet\(\frac{\tau}{z},M^2\)
\nonumber\\
&=(1-\tau)^\xi\Delta(\xi)\sum_{n=0}^\infty\frac{1}{n+\xi}
\frac{1}{n!}(1-\tau)^n\sigscet^{(n)}(\tau,M^2);
\label{sigmadef}\\
&\sigscet^{(n)}(\tau,M^2)=\frac{\partial^n}{\partial\tau^n}\sigscet(\tau,M^2),
\end{align}
up to corrections suppressed by powers of $1-\tau$,
as shown in Appendix~\ref{sec:applus}.

Equation~(\ref{tappeto}) provides the sought-for explicit comparison
of the QCD and SCET results at the level of hadronic
cross-sections. Note that the non-convolutive nature of the SCET result
implies that the generating function for the correction term is now
given by the function $\Sigma(\tau,\xi)$, which depends on the parton
luminosity, rather than by the universal function $K(z,\xi)$ Eq.~(\ref{Kdef}).

In order to understand the correction term in Eq.~(\ref{tappeto}), we
note that, with the choice of $\mu_s$ Eq.~(\ref{softhad}), we get
\beq
c^\xi\Sigma(\tau,\xi)=e^{-\gamma\xi}\Delta(\xi)\sum_{n=0}^\infty
\frac{1}{n+\xi}\frac{1}{n!}(1-\tau)^n\sigscet^{(n)}(\tau,M^2),
\label{cancelxi}
\eeq
so the dependence on $(1-\tau)^\xi$ cancels.
It follows that $\xi$ derivatives acting on $c^\xi\Sigma(\tau,\xi)$
do not induce any extra logarithmic enhancement, other than that
of $\sigscet(\tau,M^2)$ itself:
\beq
\sigqcd(\tau,M^2)=\sigscet(\tau,M^2)
+\as^2(M^2)\sum_{n=0}^\infty\frac{C_n}{n!}(1-\tau)^n\sigscet^{(n)}(\tau,M^2),
\label{tappeto2}
\eeq
where the constants $C_n$ are $\tau$-independent:
\begin{align}
&C_n=\left.\(
-\frac{A_1}{3}\beta_0\frac{\partial^3}{\partial\xi^3}
+\frac{A_2}{8}\frac{\partial^2}{\partial\xi^2}
+2\hat D_2\frac{\partial}{\partial\xi}\)
\frac{e^{-\gamma\xi}\Delta(\xi)}{n+\xi}\right|_{\xi=0};
\\
&\qquad C_0=-\frac{2}{3}\zeta_3 A_1\beta_0-\frac{\pi^2}{48}A_2,
\\
&\qquad C_n=\frac{A_1\beta_0}{n}\(\frac{\pi^2}{6}-\frac{2}{n^2}\)
-\frac{A_2}{4n^2}+\frac{2\hat D_2}{n}, \qquad n>0.
\end{align}

Therefore, up to order $\as^2$, the correction term is just 
$\as^2(M^2)$ times a linear combination of
derivatives of $\sigscet$ with respect to $\ln(1-\tau)$.
It follows that the correction term is at most
of order
\beq
\as^2\times \as^k\ln^{2k}(1-\tau)\times \ln^p(1-\tau)=
\as^{h}\ln^{2h+p-4}(1-\tau);\quad h\equiv k+2,
\label{powcount}
\eeq
where terms of order $\as^k\ln^{2k}(1-\tau)$ are due to the
coefficient functions, while terms of order $\ln^p(1-\tau)$ are due to
the parton luminosity. 

In other words, at order $\as^n$, terms $\ln^k(1-\tau)$ in the SCET
and QCD result coincide if  $2n-3+p\leq k\leq 2n$. 
There are now various possibilities. If we simply neglect all
logarithmic enhancements from the parton luminosity, i.e.\  if we set $p=0$,
then we conclude that the SCET and QCD results differ by terms which are
NNLL according to the QCD counting
Tab.~\ref{tab:QCDcount}, but N$^3$LL correction according to
the SCET counting Tab.~\ref{tab:SCETcount}. Hence we conclude that,
neglecting logarithmic enhancements from the luminosity,
the SCET result does reproduce the QCD result to NNLL accuracy, albeit
with the less accurate SCET definition of what is meant by
NNLL. However, if a logarithmic enhancement from the luminosity is
present, this is no longer the case, and the discrepancy can become
arbitrarily large (i.e.\  even at the leading log level) by just
increasing the value of $p$. Note that this in particular means that
with this choice of soft scale it is
not possible to upgrade the accuracy of the  SCET expression, as given
in Tab.~\ref{tab:SCETcount}, to that of the QCD expression, as given
in Tab.~\ref{tab:QCDcount}, because at each logarithmic order the SCET
expression differs from the perturbative QCD results by terms which,
though consistent with the accuracy of Tab.~\ref{tab:SCETcount},
spoil the higher logarithmic accuracy of Tab.~\ref{tab:QCDcount}.

One may then ask whether logarithmic enhancements due to the PDF are
expected to be present, and whether they should be counted. Because
the $P_{qq}$ and $P_{gg}$ 
splitting functions behave as $P\sim \frac{1}{(1-x)}_+$ as $x\to1$,
contributions to all parton distributions $f(x)$ which are enhanced by
$\ln(1-x)$ terms will always be induced by perturbative  evolution. In
a parton distribution evaluated at some reference scale $Q_0$ these terms
will be accompanied by powers of $\alpha_s(Q_0^2)$, which is not small
if the reference scale is taken as some low ``initial'' scale. Hence,
in general, one does expect logarithmically enhanced contributions to
PDFs, unless one wishes to make some fine-tuned assumption about the
PDF itself, which can only hold at one single scale.
The second question is whether these terms should be included or not
in the power counting Eq.~(\ref{powcount}). This is a question which
cannot be answered on the basis of first principles. Two relevant
observations here are the following. First, once one substitutes any
explicit expression of the parton luminosity in the expression
Eq.~(\ref{tappeto}) for the difference between the SCET and QCD
result, there is no way to separate what comes from the luminosity and
what comes from the coefficient function, because the SCET expression
is not factorizable. Hence, in order to discard the luminosity logs
from the power counting ones has to invoke the explicit SCET
expression Eq.~(\ref{Cscet}): in other words, one must argue that
the SCET expression contains more information than that which is
contained in the order-by-order perturbative result. The second
observation is that in practice these correction terms may be
parametrically large in realistic situations, and they may lead to significant
discrepancies between the predictions obtained using $\sigscet(\tau,M^2)$ or
$\sigqcd(\tau,M^2)$. 

\subsection{All orders}
\label{sec:allord}

The fixed $\Ord(\as^2)$ computation of Section~\ref{sec:fixord} can be
easily generalized to all orders. First of all, we note that 
the argument is based on the observation that to  $\Ord(\as^2)$ the
correction term Eq.~(\ref{tappeto}) can be expressed as a series of
derivatives of the function $c^\xi \Sigma(\tau,\xi)$ with respect to
$\xi$ , but these do not
lead to an extra logarithmic enhancement beyond that which is already
present in $\Sigma(\tau,\xi)$. The argument of Section~\ref{sec:fixord}
would thus hold, to NNLL
but to all orders in $\alpha_s$, provided only the correction term in
Eq.~(\ref{tappeto}) was a series of series of
derivatives of the function $c^\xi \Sigma(\tau,\xi)$ with respect to
$\xi$ to all orders in $\alpha_s$. This is true if and only if
$\hat{\cal S}_r$ depends on $\mu_s$ only through powers of $\ln
\frac{c}{N}$, with $c$ given by Eq.~(\ref{cdef}).
  
 Now, we  observe
that the generic term in $\hat{\cal S}_r$, Eq.~\eqref{Sr}, has the form
\beq
\int_{\mu_s^2}^{M^2/\bar{N}^2}\frac{d\mu^2}{\mu^2}\,\as^n(\mu^2)
\(\ln\frac1{\bar{N}^2}-\ln\frac{\mu^2}{M^2}\)^m
=\int_1^{c^2/N^2}\frac{dt}{t}\,\as^n(t\mu_s^2)\(\ln\frac{c^2}{N^2}-\ln t\)^m,
\eeq
with $n\ge 2$ and $m=0,1,2$. This is not a function of $\ln\frac cN$ only,
because of the dependence of $\as$ on $\mu_s$. In order to generalize
the argument to all orders we must thus study this dependence. Note that 
because
\beq
\as^n(t\mu_s^2)=\as^n(M^2)
-n\beta_0\as^{n+1}(M^2)\ln\frac{t\mu_s^2}{M^2}+\Ord(\as^{n+2})
\eeq
terms in  $\hat{\cal S}_r$ which are
not a function of $\ln\frac cN$ only first appear
at order $\as^3(M^2)$.

Furthermore,
\beq
\as^{n+1}(M^2)\ln\frac{\mu_s^2}{M^2}
\int_1^{c^2/N^2}\frac{dt}{t}\,
\(\ln\frac{c^2}{N^2}-\ln t\)^m
=\frac{1}{m+1}\as^{n+1}(M^2)\ln\frac{\mu_s^2}{M^2}
\ln^{m+1}\frac{c^2}{N^2}.
\eeq
For $n=2$, this term contributes to Eq.~\eqref{tappeto}
an order-$\as^3$ correction. This gives a series of extra
contributions  to the correction term, on top of those whose order was
given in  Eq.~(\ref{powcount}), which are at most of order
\beq
\as^3\ln(1-\tau)\times \as^k\ln^{2k}(1-\tau) \times \ln^p(1-\tau)
=\as^{h}\ln^{2h-5+p}(1-\tau);\quad h\equiv k+3
\label{powcount1}
 \eeq
while higher-order terms are even more suppressed. 

The power counting which ensues from Eq.~(\ref{powcount1}) is the same
as that of Section~\ref{sec:fixord}: neglecting logarithmic
enhancements from the luminosity (i.e.\ if $p=0$), the SCET result
does reproduce the QCD result to NNLL accuracy, but with the less
accurate SCET definition of what is meant by NNLL. If a logarithmic
enhancement from the luminosity is included, the QCD and SCET results
differ, with the discrepancy appearing at any desired logarithmic
order (including leading log) if the enhancement of the luminosity is
sufficiently strong.

It is interesting to observe that a different power counting might be
appopriate in the phenomenologically relevant case in which the
hadronic $\tau$ is actually far from threshold, yet threshold
resummation effects are non-negligible because the partonic
center-of-mass energy is lower than the hadronic one, as discussed in
the beginning of Section~\ref{sec:soft}. In this case, it might be
appropriate to simply take $\mu_s$ as some numerical constant, and
compare directly the SCET and QCD expressions of the coefficient
function $\Cscet$ and $\Cqcd$ through the master formula
Eq.~(\ref{CCCmom}).  But if $\tau$ is far from threshold, then $\mu_s$
Eq.~\eqref{softhad} is of the same order as the hard scale $M$.  As a
consequence, in this case $C_r(N,M^2,\mu_s^2)$ Eq.~\eqref{crfin}
manifestly starts at NLL order, as confirmed by inspection of
Eq.~\eqref{CrNa2}, which gives
\beq
C_r(N,M^2,\mu_s^2)=1+\Ord\(\as^2(M^2)\ln^3\frac{1}{N}\).
\eeq
One must therefore conclude that this class of NLL terms are resummed,
through the leading-log function $g_1$ in Eq.~\eqref{eq:S}, by the QCD
result $\Cqcd(z,M^2)$ but are not resummed at all in $\Cscet(z,
M^2,\mu_s^2)$.  This problem may be alleviated through generalizations
of the choice Eq.~\eqref{softhad} such as those proposed in
Ref.~\cite{Becher:2007ty} (and mentioned above at the beginning of
Section~\ref{sec:hadronic}), whereby one rescales $\mu_s$
Eq.~\eqref{softhad} by a factor (determined for instance using
scale-optimization methods), because they lead to smaller values of
$\mu_s$. These choices, however, do not affect the counting of logs,
and it is therefore impossible to assess their impact in a comparison
with standard QCD results, other than by numerical methods.


\section{Summary}
\label{sec:concl}
We have analyzed in detail the relation between the 
approach to threshold resummation based on perturbative factorization
of
Refs.~\cite{Catani:1989ne,Sterman:1986aj,Contopanagos:1996nh,Forte:2002ni},
and the SCET approach of
Refs.~\cite{Becher:2006nr,Becher:2006mr,Becher:2007ty,Ahrens:2008qu},
with the main goal of exploring the viability, both theoretical and
phenomenological,  of the SCET prescription
to treat the divergent nature of the perturbative QCD expansion in the
soft limit. By deriving 
a master formula which connects resummed results in these two
approaches, we have shown that the way they are related depends on the
choice of soft scale in the SCET expression. We have explicitly
performed calculations up to next-to-next-to-leading logarithmic
accuracy, though it is easy to convince oneself that the structure of
our master formula holds to any logarithmic order.

We have shown that if
SCET resummation is performed in Mellin space, then it coincides with
the standard perturbative result. The SCET and QCD results then have
the same accuracy, and are both beset by the problem of the divergence
of the perturbative expansion.  With this (partonic) choice of soft
scale SCET and QCD  provide alternative ways of
deriving the same resummed result. 

If SCET resummation is performed in
momentum space, as advocated in Ref.~\cite{Becher:2006nr},
the SCET and QCD results differ by a non-universal term, which depends
on the parton luminosity (explicitly given up to $\Ord(\alpha_s^2)$ in
Eq.~(\ref{CrNa2})):  the SCET approach separates off the series
of divergent contributions which is then contained in this term, with
the SCET resummed result now given by a convergent  perturbative
expansion. 
The price to pay for this is fourfold. First, because the
difference term
is non-universal, it may spoil the logarithmic accuracy of the
resummed result depending on the parton luminosity. 
In particular, if the parton luminosity contains
logarithmically enhanced contributions (as it generally will, based on
its behaviour upon QCD evolution), the difference term may enter at
any logarithmic accuracy (including at the leading-log level), {\it
  unless} one decides that logarithms coming from the luminosity
should not be included in the power counting. Note however that,
because this correction term is not factorized, there is no way of
actually isolating the logs that come from the luminosity, other than
to assume that the luminosity does not contain any.

If this problem of non-universality is neglected, the SCET and QCD
results are equivalent, however only by redefining the logarithmic
accuracy to be always by one power lower, according to the counting of
Tab.~\ref{tab:SCETcount} rather than the more accurate perturbative
QCD counting Tab.~\ref{tab:QCDcount}. Hence the second price to pay is
that the logarithmic accuracy of the SCET result in this case is
always lower by one power of log, to all orders in $\alpha_s$. Third,
while perturbative QCD resummation prescriptions such as the
minimal~\cite{Catani:1996yz} or
Borel~\cite{Forte:2006mi,Abbate:2007qv} prescription introduce
corrections to the perturbative result which are power suppressed or
more, the SCET prescription introduces a deviation which is only
logarithmically suppressed. And finally, the power counting and
suppression in the SCET result must be done at the level of the
hadronic scale $1-\tau$, while in QCD it is done at the level of the
partonic scale $1-z$. In many cases of physical
interest~\cite{Bonvini:2010tp} it may turn out that the latter is
small even when the former isn't: in these cases the QCD counting will
be more accurate.

It will be interesting to investigate the phenomenological implications
of this state of affairs. Our result enables such an investigation, by
providing a closed-form expression for the difference between the SCET
and QCD results.          

\vspace{\stretch{1}}

\noindent{\bf Acknowledgements} SF and GR are grateful to Stefano
Catani for long and fruitful conversations. GR thanks Martin Beneke
for useful discussions. MB thanks the CERN Theory Unit for hospitality
while completing this work, and Frank Tackmann for comments on a
preliminary version of this paper.  Part of this work was performed during
the GGI Workshop ``High-energy QCD after the start of the LHC'',
Florence (Italy), September 5-21, 2011.

\eject

\appendix
\section{Mellin transforms}
\label{sec:mellin}
We collect here some useful results on Mellin transforms, while
referring to Section~2
of Ref.~\cite{Forte:2002ni} and the appendices of
Refs.~\cite{Abbate:2007qv,Bonvini:2010tp} for a fuller treatment.

The Mellin transform which are necessary for the computation of
resummed terms, such as the exponent of Eq.~\eqref{Ctrad}, can be
performed using
\beq
\int_0^1 dz\,z^{N-1}\,\left[\frac{\ln^p(1-z)}{1-z}\right]_+
=
-\sum_{k=0}^{p+1}
\frac{\Gamma^{(k)}(1)}{k!}
\frac{d^k}{dL^k}\int_0^{1-1/N}dz\,\frac{\ln^p(1-z)}{1-z}
+\Ord\(\frac{1}{N}\)
\label{Mellin}
\eeq
where $L=\ln\frac{1}{N}$ Eq.~(\ref{eq:ab_def}). 

At NNLL
\begin{align}
&\int_0^1 dz\,z^{N-1}\,\left[\frac{F\(\ln(1-z)\)}{1-z}\right]_+
\nonumber\\
&\quad =-\[1-\gamma\frac{d}{dL}
+\frac{1}{2}\(\gamma^2+\frac{\pi^2}{6}\)
\frac{d^2}{dL^2}\]
\int_0^{1-1/N}dz\,\frac{F\(\ln(1-z)\)}{1-z}+\text{N$^3$LL}+\Ord\(\frac{1}{N}\),
\label{MellinNNLL}
\end{align}
(where $\gamma=-\Gamma'(1)$ is the Euler constant)
for any function $F(\ell)$ which admits a Taylor expansion around
$\ell=0$.
Now,
\begin{align}
\(1-\gamma\frac{d}{dL}+\frac{\gamma^2}{2}\frac{d^2}{dL^2}\)L^p
&=L^p-\gamma pL^{p-1}+\frac{p(p-1)}{2}\gamma^2L^{p-2}
\nonumber\\
&=\(L-\gamma\)^p+\Ord(L^{p-3}).
\label{LbarNNLL}
\end{align}
Hence, to NNLL accuracy, Eq.~\eqref{Mellin} can be written in the equivalent 
form
\begin{align}
\int_0^1 dz\,z^{N-1}\,&\left[\frac{F\(\ln(1-z)\)}{1-z}\right]_+
\nonumber\\
&=-\int_0^{1-1/\bar{N}}dz\,\frac{F\(\ln(1-z)\)}{1-z}
-\frac{\pi^2}{12}
\frac{d^2}{dL^2}\int_0^{1-1/N}dz\,\frac{F\(\ln(1-z)\)}{1-z}
\label{MellinNNLL2}
\end{align}
where
\beq
\bar{N}=Ne^\gamma;\qquad\ln\frac{1}{\bar{N}}=L-\gamma.
\label{Lbar}
\eeq

An essential ingredient in the discussion of Section~\ref{sec:hadronic}
is the inverse Mellin transform of
\beq
\ln^n\frac{c}{N}
\eeq
where $c$ is a constant. In order to compute it, we start from the identity
\beq
\ln^n\frac{1}{N}
=\left.\frac{d^n}{d\xi^n}\Delta(\xi)
\int_0^1dz\,z^{N-1}\ln^{\xi-1}\frac{1}{z}\right|_{\xi=0},
\label{MellinlogN}
\eeq
where $\Delta(\xi)=\frac{1}{\Gamma(\xi)}$. 
Eq.~\eqref{MellinlogN} should be (and usually is) written
in the form
\beq
\ln^n\frac{1}{N}
=\left.\frac{d^n}{d\xi^n}\Delta(\xi)
\int_0^1dz\,z^{N-1}\[\ln^{\xi-1}\frac{1}{z}\]_+\right|_{\xi=0}
+\delta_{n0},
\label{MellinlogNg}
\eeq
so that the integral is well defined
even when the derivatives and the limit $\xi\to 0$ are taken
under the integral sign. This is not necessary for our present purposes.
Rewriting Eq.~\eqref{MellinlogN} with $N$ replaced by $N/c$ we obtain
\begin{align}
\ln^n\frac{c}{N}
&=\left.\frac{d^n}{d\xi^n}\Delta(\xi)\int_0^1dz\,z^{\frac{N}{c}-1}
\ln^{\xi-1}\frac{1}{z}\right|_{\xi=0}
\nonumber\\
&=\left.\frac{d^n}{d\xi^n}c^\xi\Delta(\xi)\int_0^1dz\,z^{N-1}
\ln^{\xi-1}\frac{1}{z}\right|_{\xi=0}
\label{MellinlogcN}
\end{align}
after rescaling the integration variable $z\to z^c$.
The inverse Mellin transform of $\ln^n\frac{c}{N}$ can now be  immediately
read off Eq.~\eqref{MellinlogcN}.

\section{Convolutions}
\label{sec:applus}
In this Appendix we compute the integral
\beq
\Sigma(\tau,\xi)=
\Delta(\xi)\int_\tau^1\frac{dz}{z}\,\sigma\(\frac{\tau}{z}\)
\ln^{\xi-1}\frac{1}{z},
\eeq
where $\sigma(\tau)\equiv\sigscet(\tau,M^2)$ for simplicity,
up to terms suppressed by powers of $1-\tau$.
Using
\beq
\ln\frac{1}{z}=1-z+\Ord((1-z)^2)
\eeq
we find
\begin{align}
\Sigma(\tau,\xi)&=
\Delta(\xi)\int_\tau^1\frac{dz}{z}\,(1-z)^{\xi-1}\sigma\(\frac{\tau}{z}\)
\nonumber\\
&=\Delta(\xi)\sum_{n=0}^\infty\frac{\sigma^{(n)}(\tau)}{n!}\tau^n
\int_\tau^1dz\,(1-z)^{\xi-1}\frac{(1-z)^n}{z^{n+1}}.
\label{Sigma1}
\end{align}
Expanding $1/z^{n+1}$ in powers of $1-z$ we see that the integral
is a sum of terms proportional to
\beq
(1-\tau)^{\xi+m};\qquad m\geq n.
\eeq
Hence, the derivatives $\sigma^{(n)}(\tau)$ appear in $\Sigma(\tau,\xi)$
multiplied by $(1-\tau)^m$, with $m\ge n$. Now
\begin{align}
(1-\tau)\sigma^{(1)}(\tau)&=-\frac{d\sigma(\tau)}{d\ln(1-\tau)}
\nonumber\\
(1-\tau)^2\sigma^{(2)}(\tau)&=
-\frac{d\sigma(\tau)}{d\ln(1-\tau)}
+\frac{d^2\sigma(\tau)}{d\ln^2(1-\tau)}
\nonumber\\
&\ldots
\end{align}
which means that $(1-\tau)^m\sigma^{(n)}(\tau)$ with $m>n$ is power-suppressed,
and can be we neglected in Eq.~\eqref{Sigma1} since we are only interested
in logarithmically-enhanced contributions to $\Sigma(\tau,\xi)$. Hence
\beq
\Sigma(\tau,\xi)=\Delta(\xi)(1-\tau)^\xi\sum_{n=0}^\infty
\frac{(1-\tau)^n\sigma^{(n)}(\tau)}{n!(n+\xi)}
+\text{power-suppressed terms}.
\eeq

\end{document}